\documentclass[prd,onecolumn,amsmath,amssymb,superscriptaddress,nofootinbib,12pt]{revtex4-2}
\usepackage{url}

\usepackage{epsfig}
\usepackage{amsfonts}
\usepackage{graphicx}
\usepackage{subfigure}
\usepackage{epsfig}
\usepackage{eepic}
\usepackage{amsmath}
\usepackage{amssymb}
\usepackage{color}
\usepackage{bbm}
\usepackage{dcolumn}
\usepackage{bm}
\usepackage[normalem]{ulem}
\usepackage{mathrsfs}
\usepackage{bbold}
\usepackage{datetime}

\usepackage{overpic} 
\usepackage{rotating}
\usepackage[usenames,dvipsnames]{xcolor}
\usepackage[colorlinks=true,citecolor=Magenta,linkcolor=Green,urlcolor=Green]{hyperref}
\usepackage{lipsum} 
\usepackage{tikz,tikz-3dplot}
\usetikzlibrary{shapes.geometric}
\usepackage{etoolbox} 
\usepackage[capitalize]{cleveref}
\usepackage{extarrows}
\usepackage{ragged2e}

\def\bc{\begin{center}}

\def\ec{\end{center}}
\def\be{\begin{eqnarray}}
\def\ee{\end{eqnarray}}
\definecolor{dyellow}{rgb}{1.,0.8,.0}
\definecolor{myblue}{rgb}{.1,.1,.7}
\definecolor{dcyan}{rgb}{.0,.6,.6}
\definecolor{dmagenta}{rgb}{0.6,0.0,0.6}
\definecolor{brown}{rgb}{0.6,0.2,0.}
\definecolor{darkblue}{rgb}{.0,.0,0.5}
\definecolor{darkred}{rgb}{0.75,0.0,0.0}
\definecolor{orange}{rgb}{1.,.6,.0}
\definecolor{dorange}{rgb}{0.8,.4,.0}
\definecolor{darkgreen}{rgb}{0.0,0.6,0.0}
\definecolor{purple}{rgb}{.4,.0,.4}
\definecolor{lightgrey}{rgb}{0.7, 0.7, 0.7}
\definecolor{grey}{rgb}{0.4, 0.4, 0.4}

\usepackage{geometry}
\usepackage[font=small,labelfont=bf,justification=raggedright]{caption}

\newcommand{\xdownarrow}[1]{%
  {\left\downarrow\vbox to #1{}\right.\kern-\nulldelimiterspace}
}
\newcommand{\xuparrow}[1]{%
  {\left\uparrow\vbox to #1{}\right.\kern-\nulldelimiterspace}
}

\definecolor{myred}{RGB}{189, 38, 49}

\begin{document}
\title{Holographic Timelike Entanglement Entropy from Rindler Method}
\author{Peng-Zhang He}\email{hepzh@buaa.edu.cn}
\affiliation{Center for Gravitational Physics, Department of Space Science, Beihang University, Beijing 100191, China}
\author{Hai-Qing Zhang} \email{hqzhang@buaa.edu.cn}
\affiliation{Center for Gravitational Physics, Department of Space Science, Beihang University, Beijing 100191, China}
\affiliation{Peng Huanwu Collaborative Center for Research and Education, Beihang University, Beijing 100191, China}

	\begin{abstract}
		For a Lorentzian invariant theory, the entanglement entropy should be a function of the  domain of dependence of the subregion under consideration. More precisely, it should be a function of the  domain of dependence and the appropriate cut-off. In this paper, we refine the concept of cut-off to make it applicable to timelike regions and assume that the usual entanglement entropy formula also applies to timelike intervals. Using the Rindler method, the timelike entanglement entropy can be regarded as the thermal entropy of the CFT after the Rindler transformation plus a constant $ic\pi/6$ with $c$ the central charge. The gravitational dual of the `{\it covariant}' timelike entanglement entropy is finally presented following this method.
	\end{abstract}
	
	
	\maketitle

	\section{Introduction} 
	The Anti-de Sitter/Conformal Field Theory (AdS/CFT) correspondence \citep{Maldacena:1997re,Gubser:1998bc,Witten:1998qj} provides an interesting perspective on gravity and CFT: quantum gravity in $(d+1)$-dimensional AdS spacetime is equivalent to a CFT on the $d$-dimensional boundary. Over the past two decades, increasing evidences have supported this correspondence \cite{Kovtun:2004de,Hartnoll:2008vx,Lee:2008xf,Cubrovic:2009ye,Liu:2009dm,Zaanen:2015oix}. Among these, the holographic entanglement entropy \citep{Ryu:2006bv,Ryu:2006ef,Hubeny:2007xt} is one of the strongest supports for it.
	In a quantum theory, if we divide the entire Hilbert space into two parts $A$ and $B$, the entanglement entropy $S_A$ is defined as the von Neumann entropy of the reduced density matrix $\rho_A$ 
	\begin{equation}
		S_A=-\mathrm{Tr}_A\left( \rho _A\log \rho _A \right), \label{ee}
	\end{equation}
	where $\rho_A=\mathrm{Tr}_B(\rho)$ with $\rho$ the density matrix for the whole system.	 In general, the calculation of entanglement entropy in CFT is not easy \citep{Calabrese:2005zw,Calabrese:2004eu,Calabrese:2009qy}. Fortunately, Ryu and Takayanagi proposed a method to calculate the entanglement entropy of a given region in CFT utilizing the AdS/CFT correspondence \citep{Ryu:2006bv}. The Ryu-Takayanagi formula is very elegant and simple. In short, the entanglement entropy of a region $A$ in the boundary CFT can be given by the area of the minimal surface $m_A$ in the bulk that having the same boundary as $A$ (denoted as $\partial A$),
	\begin{equation}
		S_{A}=\frac{1}{4 G_{N}} \operatorname{Area}\left(m_{A}\right),
	\end{equation}
	where $G_N$ is the Newton's constant. 
	
	Since the initial proposal by Ryu and Takayanagi, there are several attempts trying to prove the Ryu-Takayanagi formula \citep{fursaev2006proof,Casini:2011kv,lewkowycz2013generalized}. Among these, reference \citep{Casini:2011kv} was the first to use the conformal transformation to prove the formula in the case of a spherical entanglement surface (i.e., $\partial A$). The basic idea of the proof is to utilize the conformal transformations to convert the calculation of the entanglement entropy into the calculation of the thermal entropy in the CFT, which can then be mapped into the calculation of the black hole entropy in the gravitational dual according to the AdS/CFT correspondence. This method was later developed to derive the holographic entanglement entropy formula in other holographic theories \citep{Jiang_2017,Song:2016gtd,Castro:2015csg} and is known as the ``Rindler method". 
	
	There are many kinds of entanglement entropy \cite{Nielsen:2012yss}.  Among these, pseudo-entropy is an interesting generalization of the entanglement entropy \citep{Nakata:2020luh,Guo:2022sfl,Guo:2022jzs,He:2023wko}. Its definition is similar to that of the entanglement entropy,
	\begin{equation}
		S_A=-\mathrm{Tr}\left[ \tau _A\log \tau _A \right] ,\label{pse}
	\end{equation}
	where $\tau_A$ is given by two pure states $\left|\psi\right>$ and $\left|\varphi\right >$,
	\begin{equation}
		\tau _A=\mathrm{Tr}_B\left[ \frac{\left| \psi \right> \left< \varphi \right|}{\left< \varphi \middle| \psi \right>} \right] .\label{tau}
	\end{equation}
	Although the forms of \eqref{pse} and \eqref{ee} are similar, pseudo-entropy is usually complex  since Eq.\eqref{tau} is not Hermitian. When $\left |\psi\right >=\left |\varphi\right >$, pseudo-entropy becomes ordinary entanglement entropy as defined in Eq.\eqref{ee}. 
	When calculating the holographic entanglement entropy, the subregions considered are spacelike regions on the boundary. However, recent work \citep{Wang:2018jva} has shown that the traditional spacelike entanglement entropy does not fully capture the entangling properties of CFTs and timelike entanglement entropy needs to be introduced. Interestingly, later study \citep{Doi:2022iyj} found that the pseudo-entropy in dS/CFT is directly related to the timelike entanglement entropy in AdS/CFT. This timelike entanglement entropy is defined by analytically continuing a spacelike subregion into a timelike one. Further, the paper \citep{Doi:2023zaf} suggests that the time coordinate may emerge from the imaginary part of the timelike entanglement entropy or pseudo-entropy, generalizing the familiar idea that the space coordinate can emerge from the ordinary entanglement entropy \cite{VanRaamsdonk:2009ar,Swingle:2009bg,Swingle:2012wq,Swingle:2017blx}. Therefore, timelike entanglement entropy appears to be an important generalization of the ordinary entanglement entropy. Related work on the timelike entanglement entropy can be found in references \citep{Narayan:2022afv,Reddy:2022zgu,Li:2022tsv,Doi:2023zaf,Jiang:2023ffu,Narayan:2023ebn,Jiang:2023loq,Chu:2023zah}. 
	
The timelike entanglement entropy is currently considered as a special pseudo-entropy. In the AdS/CFT framework, both the real and imaginary parts of pseudo-entropy have clear spacetime geometric interpretations \cite{Nakata:2020luh}. Moreover, studies in quantum many-body systems suggest that pseudo-entropy can be used to detect quantum chaos in the system \cite{Goto:2021kln} and to distinguish between different quantum phases \cite{Mollabashi:2020yie}. Pseudo-entropy has been widely discussed in holographic duality, quantum field theory, quantum information, and quantum many-body systems \cite{Nakata:2020luh,Doi:2022iyj,Guo:2022sfl,Fullwood:2023qhw,Mollabashi:2021xsd}. However, from the worldwide, the physical significance of timelike entanglement entropy is not yet well understood.
	
	In this paper, we want to use the Rindler method to re-examine the (holographic) timelike entanglement entropy in AdS$_3$/CFT$_2$ correspondence.  Because the introduction of the cut-off when calculating entanglement entropy using the replica trick \citep{Calabrese:2009qy,Calabrese:2005zw,Calabrese:2004eu} in field theory seems somewhat arbitrary, our treatment of the cut-off is different from those in previous works \citep{Doi:2022iyj,Doi:2023zaf}. Specifically, our idea is that the entanglement entropy should be generally a function of the domain of dependence $\mathcal{D}$ and the cut-off of the subregion under consideration. When calculating entanglement entropy, a spacelike cut-off $\varepsilon$ is always introduced. However, if we extend the entanglement entropy to timelike subregions, the cut-off naturally becomes timelike. We use this timelike cut-off $ \varepsilon^{t}$ to replace the previous spacelike cut-off, and consequently the timelike entanglement entropy  $S^{\left( t \right)}$ becomes
	\begin{equation}\label{stimelike}
		S=S\left( \mathcal{D} ,\varepsilon \right) \rightarrow S^{\left( t \right)}\equiv S\left( \mathcal{D} ,\varepsilon ^t \right) ,
	\end{equation}
	which is exactly the timelike entanglement entropy studied in references \citep{Doi:2022iyj,Doi:2023zaf}.
	
	The structure of this article is arranged as follows: In Sec.\ref{sec2}, we will briefly review the previous definition of timelike entanglement entropy and its holographic dual in AdS$_3$/CFT$_2$. In Sec.\ref{sec3}, we will first review how to calculate holographic entanglement entropy using the Rindler method. Then we will give the definition of timelike entanglement entropy and use the Rindler method to give the gravitational dual of the timelike entanglement entropy. Finally, we draw the conclusions and discussions in Sec.\ref{sec:con}.

	\section{Brief review of timelike entanglement entropy}\label{sec2}
	
	
	Consider a spacelike interval $A=[(t_1,x_1),(t_2,x_2)]$ in a $(1+1)$ dimensional Minkowski space (with the signature of the metric as $(-1,+1)$), and define
	\begin{equation}
		T_0\equiv t_2-t_1,\qquad X_0\equiv x_2-x_1.
	\end{equation}
	Using the replica trick \citep{Calabrese:2009qy,Calabrese:2005zw,Calabrese:2004eu}, one can obtain the entanglement entropy of the subregion $A$ as
	\begin{equation}
		S_A=\frac{c}{3}\log \frac{\sqrt{X_{0}^{2}-T_{0}^{2}}}{\varepsilon},
	\end{equation}
	where $c$ is the central charge of the CFT and $\varepsilon$ is a UV cut-off. Now suppose that the entanglement entropy formula also applies to the timelike interval, which defines the timelike entanglement entropy \citep{Doi:2023zaf}. Thus, $X^2_0-T^2_0<0$. Consequently, we get the timelike entanglement entropy
	\begin{equation}
		S_{A}^{\left( t \right)}=\frac{c}{3}\log \frac{\sqrt{T_{0}^{2}-X_{0}^{2}}}{\varepsilon}+\frac{ic\pi}{6}.\label{2.3}
	\end{equation}
	The definition of timelike entanglement entropy can also be extended to finite size CFT and finite temperature CFT, respectively, 
	\begin{eqnarray}
		S_{R}^{\left( t \right)}&=&\frac{c}{6}\log \left[ \frac{R^2}{\pi ^2\varepsilon ^2}\sin \left( \frac{\pi}{R}\left( \Delta t+\Delta \phi \right) \right) \sin \left( \frac{\pi}{R}\left( \Delta t-\Delta \phi \right) \right) \right] +\frac{ic\pi}{6},\\
		S_{\beta}^{\left( t \right)}&=&\frac{c}{6}\log \left[ \frac{\beta ^2}{\pi ^2\varepsilon ^2}\sinh \left( \frac{\pi}{\beta}\left( \Delta t+\Delta x \right) \right) \sinh \left( \frac{\pi}{\beta}\left( \Delta t-\Delta x \right) \right) \right] +\frac{ic\pi}{6},
	\end{eqnarray}
	where $R$ is the total length of the finite system, $\Delta t$ is the difference between the time coordinates of the endpoints of the considered interval, $\Delta \phi$ and $\Delta x$ are the differences between the coordinates in the spatial direction of the considered interval, $\beta$ is the inverse of the finite temperature.
	
	
	\begin{figure}[t]
		\centering
		\includegraphics[width=0.25\linewidth]{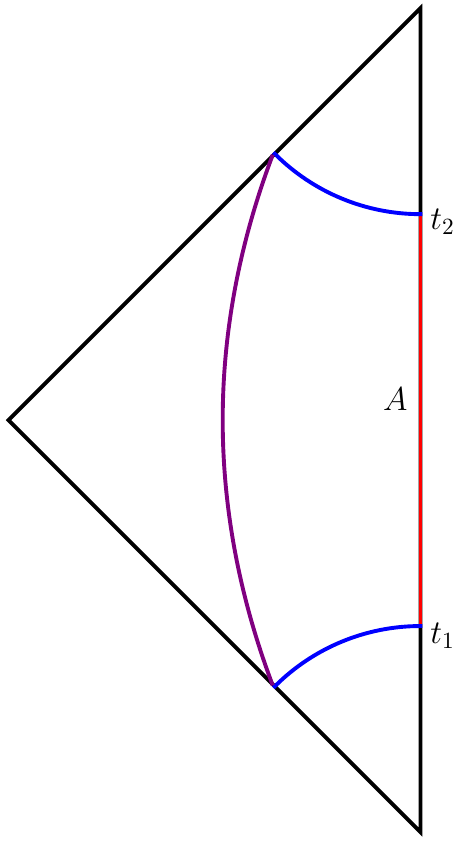}
		\caption{Penrose diagrams of the geodesics connecting the boundary subregions, where the boundary subregion $A$ is denoted as the red line segment. The blue curves represent the spacelike geodesics, while the purple curve represents the timelike geodesic. }
		\label{fig:f2}
	\end{figure}
	
	In terms of the AdS$_3$/CFT$_2$ correspondence, one can define the metric of the AdS$_3$ spacetime in Poincar\'e coordinate as (The AdS radius is set to $\ell_{\rm{AdS}}=1$),
	\begin{equation}
		d s^{2}=\frac{d z^{2}-d t^{2}+d x^{2}}{z^{2}}.
	\end{equation}
	In the boundary CFT$_2$, we consider a timelike interval of length $T_0$ with fixed spatial coordinates. In this case, there are no spacelike geodesics which can directly connect the boundaries of the timelike interval in the CFT. However, this timelike interval can be connected by two spacelike geodesics and one timelike geodesic in the bulk of AdS$_3$ as shown in Fig. \ref{fig:f2}. The spacelike geodesics (in blue) connect the endpoints of the interval to null infinity, while the timelike geodesic (in purple) connects the endpoints of the two spacelike curves at the null infinity \citep{Doi:2022iyj}. Therefore, the spacelike geodesics are given by 
	\begin{equation}
		t= \pm \sqrt{z^{2}+T_{0}^{2} / 4}, \quad z \in(0, \infty).\label{77}
	\end{equation}
	and the form of the timelike geodesic is
	\begin{equation}
		z=\sqrt{\left( t-t_0 \right) ^2+z_{0}^{2}},\qquad t\in \left( -\infty ,\infty \right) ,\label{88}
	\end{equation}
	where $t_0$ and $z_0$ are two constants determined by the endpoint positions of the spacelike geodesic at the null infinity. Final results indicate that \textit{two spacelike geodesics contribute to the real part of the timelike entanglement entropy, while the timelike geodesic gives the imaginary part of the timelike entanglement entropy}.

	It is worth noting that the appearance of the imaginary part in the definition of the timelike entanglement entropy seems very strange no matter how you look at it. There seems to be an analytic continuation to make the holographic entanglement entropy applicable to timelike intervals, but the cut-off is still the cut-off corresponding to spacelike intervals, therefore an imaginary part appears. In this way, there can naturally be an equivalent definition, where the interval is still a spacelike interval, but the cut-off is taken in the timelike interval. This is precisely the idea behind another way of defining timelike entanglement entropy in reference \citep{Doi:2023zaf}. (We briefly review this definition in the Appendix \ref{A}.)

	\section{Alternative perspective on timelike entanglement entropy: Rindler method}\label{sec3}
	In this section, we will give an alternative perspective on timelike entanglement entropy based on the Rindler method. For the convenience of the reader and the subsequent narrative, we will first briefly review how to compute the holographic entanglement entropy in AdS$_3$/CFT$_2$ using the Rindler method. Interested readers can refer to the references \citep{Jiang_2017,Song:2016gtd,Rangamani:2016dms,Castro:2015csg,Casini:2011kv,Wen:2018whg,He:2023cju}.
	
	\subsection{Deriving holographic entanglement entropy using Rindler method}\label{3.1}
	
	Consider a spatial interval $\mathcal{I}$, whose  domain of dependence in two dimensional Minkowski spacetime is given by
	\begin{equation}
		\mathcal{D} =\left\{ \left( u,v \right) |-\frac{l_u}{2}\le u\le \frac{l_u}{2},-\frac{l_v}{2}\le v\le \frac{l_v}{2} \right\} ,\label{99}
	\end{equation}
	in which we have set $u=x+t$ and $v=x-t$. The key to the Rindler method is to find a Rindler transformation that maps $\mathcal{D}$ to an infinitely large region $\mathcal{B}$. This transformation is a conformal transformation that maps the vacuum state in $\mathcal{D}$ to $\mathcal{B}$ and correspondingly maps the entanglement entropy of $\mathcal{I}$ to the thermal entropy in $\mathcal{B}$. For a two-dimensional CFT, the Rindler transformation is
	\begin{equation}
		\begin{aligned}
			u^{\prime} = \operatorname{arctanh} \frac{2 u}{l_{u}}, ~~~
			v^{\prime}  =\operatorname{arctanh} \frac{2 v}{l_{v}} .
		\end{aligned}
	\end{equation}
	This transformation implies that the thermal circle in $\mathcal{B}$ is 
	\begin{equation}
		\left( u^{\prime},v^{\prime} \right) \sim \left( u^{\prime}+i\pi ,v^{\prime}-i\pi \right) .
	\end{equation}
	Consequently, the thermal entropy in $\mathcal{B}$ gives rise to the desired entanglement entropy.
	
\begin{figure}[h]
		\centering
		\includegraphics[width=0.5\linewidth]{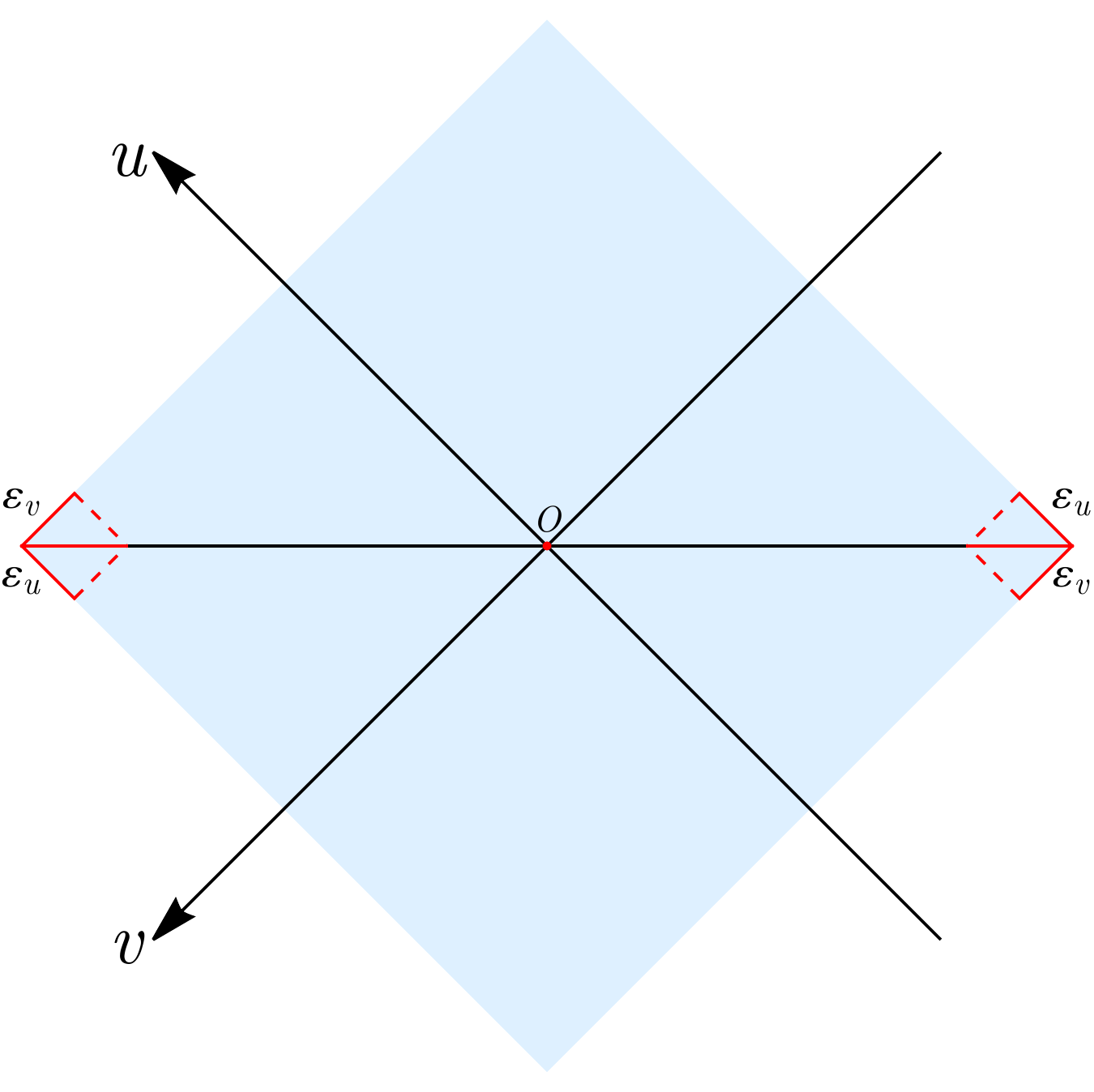}
		\caption{A schematic diagram of a spatial interval (horizontal line) and its  domain of dependence (blue shape), with cut-offs $\varepsilon_{u},\varepsilon_{v}$ marked in the diagram. }
		\label{fig:f1}
	\end{figure}

	According to the AdS/CFT correspondence, the computation of thermal entropy can be transformed into the computation of black hole entropy in the gravitational dual, which can be obtained by applying the Rindler transformation in the bulk
	\begin{eqnarray}
		z^{\prime}&=&\frac{l_{u}^{2}\left( l_{v}^{2}-4v^2 \right) +4\left( -l_{v}^{2}u^2+4\left( uv+z^2 \right) ^2 \right)}{8l_ul_vz^2},\label{4}
		\\
		u^{\prime}&=&\frac{1}{4}\log \frac{l_{v}^{2}\left( l_u+2u \right) ^2-4\left( l_uv+2\left( uv+z^2 \right) \right) ^2}{l_{v}^{2}\left( l_u-2u \right) ^2-4\left( l_uv-2\left( uv+z^2 \right) \right) ^2},\label{5}
		\\
		v^{\prime}&=&\frac{1}{4}\log \frac{l_{u}^{2}\left( l_v+2v \right) ^2-4\left( l_vu+2\left( uv+z^2 \right) \right) ^2}{l_{u}^{2}\left( l_v-2v \right) ^2-4\left( l_vu-2\left( uv+z^2 \right) \right) ^2},\label{6}
	\end{eqnarray}
	where $u, v$ and $z$ are coordinates in the Poincar\'e AdS$_3$ metric $d s^{2}=\frac{1}{z^{2}}\left(d z^{2}+d u d v\right)$. After the transformation, the new metric becomes 
	\begin{equation}
		ds^2=du^{\prime2}+dv^{\prime2}+2z^{\prime}du^{\prime}dv^{\prime}+\frac{1}{4\left( {z^{\prime}}^2-1 \right)}dz^{\prime2}.\label{7}
	\end{equation}
	Its event horizon is located at
	\begin{equation}
		z^\prime_h=1.\label{8}
	\end{equation}
	Substituting equation \eqref{8} into equation \eqref{4} yields
	\begin{equation}
		z=\frac{1}{2}\sqrt{\left( l_u+2u \right) \left( l_v-2v \right)},\label{9}
	\end{equation}
	or 
	\begin{equation}
		z=\frac{1}{2}\sqrt{\left( l_u-2u \right) \left( l_v+2v \right)}.\label{10}
	\end{equation}
	Combining equations \eqref{9} and \eqref{10} yields the RT/HRT surface
	\begin{equation}
		u=\frac{l_u v}{l_v},\qquad z=\frac12\sqrt{\left (l_v+2v\right )\left (l_u-\frac{2l_u v}{l_v}\right )}.
	\end{equation}
	Therefore, the final result of the entanglement entropy is
	\begin{equation}
		S=\frac{c}{6}\log \frac{l_ul_v}{\varepsilon _u\varepsilon _v},\label{12}
	\end{equation}
	where $\varepsilon_u$ and $\varepsilon_v$ are introduced as cutoffs in the $u$ and $v$ directions, see Fig. \ref{fig:f1}. 
	
	More generally, we can consider subsystems (i.e., spacelike curves with a  domain of dependence of $\mathcal{D}$) with different truncations at the left ends ($\varepsilon_{u1},\varepsilon_{v1}$) and right ends ($\varepsilon_{u2},\varepsilon_{v2}$). In this case, the entanglement entropy becomes
	\begin{equation}
		S=\frac{c}{12}\log \frac{l_ul_v}{\varepsilon _{u1}\varepsilon _{v1}}+\frac{c}{12}\log \frac{l_ul_v}{\varepsilon _{u2}\varepsilon _{v2}}.\label{13}
	\end{equation}
	Since $l_{u}$ and $l_{v}$ are related to the  domain of dependence of the subsystem, we see that the entanglement entropy can be eventually obtained once the subsystem is given and the truncations are chosen.
	
	\subsection{Timelike entanglement entropy from a timelike curve}
	
	As mentioned above, Eqs.\eqref{12} and \eqref{13} can be regarded as functions depending on the domain of dependence and cut-off, where the domain of dependence is characterized by $l_{u}l_{v}$. Now we want to promote the concept of entanglement entropy so that it is applicable not only to spacelike subregions but also to timelike subregions.  A natural idea is that the ``entanglement entropy" of a timelike region is also related to a spacetime subregion and cut-offs. We hope that this spacetime subregion remains the causal domain of some spacelike interval. The current issue is how to establish a connection between a spacelike interval and a timelike interval. Below, we will proceed to accomplish this task.
	
\begin{figure}[t]
		\centering
		\includegraphics[width=0.5\linewidth]{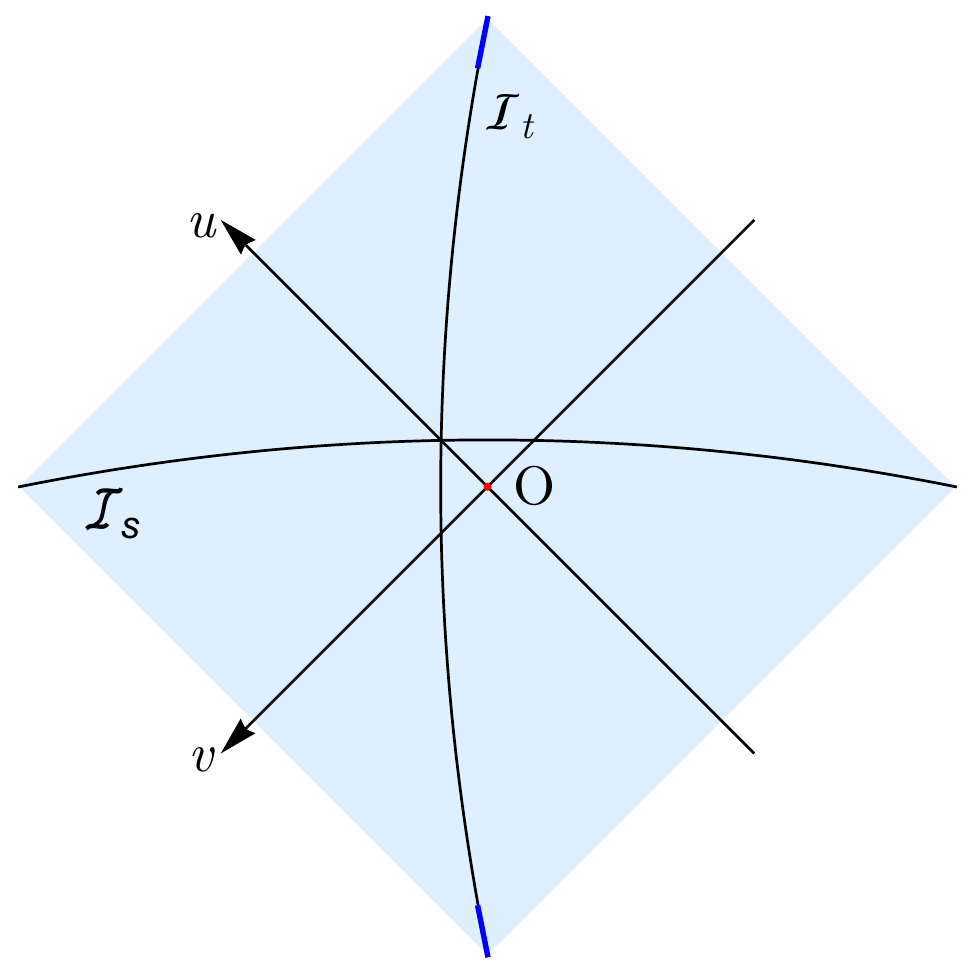}
		\caption{Illustration of the spacelike curve $\mathcal{I}_s$ and its dual timelike curve $\mathcal{I}_t$. The cut-offs of $\mathcal{I}_{t}$ are marked by the blue line segment.}
		\label{fig:f3}
	\end{figure}

	Suppose there is a spacelike curve $\mathcal{I}_s$ with parametric equations:
	\begin{equation}
		u=u(s),\qquad v=v(s),\qquad s\in [0,1],
	\end{equation}
	where $s$ is the parameter. Then we can define a timelike curve $\mathcal{I}_t$ as,
	\begin{equation}\label{uvs}
		u=u(s),\qquad v=-v(s),\qquad s\in [0,1],
	\end{equation}
	which is called the {\textit{dual}} to $\mathcal{I}_s$. Obviously, from the above relation \eqref{uvs} we can clearly see that $\mathcal{I}_s$ and $\mathcal{I}_t$ are symmetric to each other with $u$ axis as the symmetry axis. Therefore, $\mathcal{I}_s$ connects the left and right endpoints of  the domain of dependence, while $\mathcal{I}_t$ connects the upper and lower endpoints of the domain of dependence. Fig. \ref{fig:f3} provides an intuitive illustration of this setup. 
	
	Next, we define the ``cut-off" $\varepsilon_u$ and $\varepsilon_v$  on the spacelike curve with $s_1$ and $s_2$ as two parameters,
	\begin{equation}
		\varepsilon _u\equiv u\left( s_2 \right) -u\left( s_1 \right) ,\qquad \varepsilon _v\equiv v\left( s_2 \right) -v\left( s_1 \right) ,\qquad s_1,s_2\in \left[ 0,1 \right] ,
	\end{equation}
	where $s_{2}-s_{1}\rightarrow0^{+}$. In order to clarify it, let's take a specific example. Suppose there is a spacelike curve
	\begin{equation}
		\mathcal{I} _s\equiv \left\{ \left( u,v \right) |u=l_u\left( s-\frac{1}{2} \right) ,v=l_v\left( s-\frac{1}{2} \right) ,s\in \left[ 0,1 \right] \right\} .\label{17}
	\end{equation}
	After the cut-off, its regulated version is
	\begin{equation}
		\mathcal{I} _{s}^{\mathrm{reg}}\equiv \left\{ \left( u,v \right) |u=l_u\left( s-\frac{1}{2} \right) ,v=l_v\left( s-\frac{1}{2} \right) ,s\in \left[ \varepsilon ,1-\varepsilon \right] \right\} .\label{18}
	\end{equation}
	where $\varepsilon$  is a positive infinitesimal parameter whose relationship with $\varepsilon_u$ and $\varepsilon_v$ are
	\begin{equation}
		\varepsilon _u=l_u\varepsilon ,\qquad  \varepsilon _v=l_v\varepsilon .
	\end{equation}
	Therefore, we can obtain the entanglement entropy for the above subregion, which is exactly the equation Eq.\eqref{12}. 
	
	On the other hand, it is obvious that for the {\it dual} timelike versions of \eqref{17} and \eqref{18}, the truncations are
	\begin{equation}
		\varepsilon _{u}^{t}=\varepsilon _u,\qquad \varepsilon _{v}^{t}=-\varepsilon _v.
	\end{equation}
	in which the superscript $t$ denotes ``timelike". This relationship holds for any two dual curves. If we generalize the entanglement entropy in a way that Eq.\eqref{12} holds for the spacelike subregions also holds for the timelike subregions, then the entanglement entropy of the timelike subregion $\mathcal{I}_t$, which is dual to the spacelike subregion $\mathcal{I}_s$, becomes
	\begin{equation}
		S^{(t)}=\frac{c}{6}\log \frac{l_ul_v}{\varepsilon _u(-\varepsilon _v)}=\frac{c}{6}\log \frac{l_ul_v}{\varepsilon _u\varepsilon _v}+\frac{ic\pi}{6}.\label{21}
	\end{equation}
	It is evident that the entanglement entropy of the resulting timelike interval only differs from the entanglement entropy of the corresponding spacelike interval by an additional term of $ic\pi/6$.	 When $l_{u}=l_{v}$, Eq.\eqref{21} is exactly the timelike entanglement entropy introduced in reference \citep{Doi:2023zaf,Doi:2022iyj,Li:2022tsv}. More generally, we can extend Eq.\eqref{13} to the entanglement entropy of timelike subregions
	\begin{equation}
		S^{(t)}=\frac{c}{12}\log \frac{l_ul_v}{\varepsilon _{u1}\varepsilon _{v1}}+\frac{c}{12}\log \frac{l_ul_v}{\varepsilon _{u2}\varepsilon _{v2}}+\frac{ic\pi}{6}.\label{30}
	\end{equation}
	Compared to references \citep{Doi:2023zaf,Doi:2022iyj,Li:2022tsv}, the formula for timelike entanglement entropy derived here has a broader applicability. It is applicable to any timelike interval, not limited to pure null intervals. Our method of defining timelike entanglement entropy can be directly applied to the finite size CFT and finite temperature CFT, with an extra $ic\pi/6$ added, similar to Eq.\eqref{30}. We will not elaborate on it in this paper.
	
	\subsection{Holographic timelike entanglement entropy from the Rindler method}
	
	A question arising is whether the Rindler method can also be applied to derive the gravitational dual of the timelike entanglement entropy $S_{\mathrm{TEE}}$. The formal expression of \eqref{21} is
	\begin{equation}
		S_{\mathrm{TEE}}=S_{\mathrm{thermal}}+\frac{ic\pi}{6},\label{31}
	\end{equation}
	i.e., the timelike entanglement entropy is equal to the sum of the thermal entropy $S_{\mathrm{thermal}}$ undergoing a Rindler transformation  and the constant $ic\pi/6$. However, Sec. \ref{3.1} shows that the thermal entropy here corresponds to the black hole entropy, and the black hole horizon corresponds to the RT/HRT surface. Just like the usual calculation of holographic entanglement entropy, black hole entropy can only provide a real area \citep{Casini:2011kv} and cannot give an imaginary part. Therefore, it seems that this method does not directly provide a gravitational dual for timelike entanglement entropy.
	
	Nevertheless, an alternative solution is to modify the Rindler transformation so that the black hole horizon precisely corresponds to the two spacelike geodesics in the gravitational dual of timelike entanglement entropy. Mathematically speaking, there is no difference between the negative sign appearing on $\varepsilon_v$ or on $l_v$ in Eq.\eqref{21}, that is,
	\begin{equation}
		S^{\left( t \right)}=\frac{c}{6}\log \frac{l_ul_v}{\varepsilon _u\left( -\varepsilon _v \right)}=\frac{c}{6}\log \frac{l_u\left( -l_v \right)}{\varepsilon _u\varepsilon _v}.
	\end{equation}
	Following this spirit we can replace all $l_v$ in the Rindler method with $-l_v$, therefore, the previous Rindler transformation Eqs.\eqref{4}-\eqref{6} in the bulk becomes
	\begin{eqnarray}
		z^{\prime}&=&\frac{l_{u}^{2}\left( l_{v}^{2}-4v^2 \right) +4\left( -l_{v}^{2}u^2+4\left( uv+z^2 \right) ^2 \right)}{8l_u\left( -l_v \right) z^2},\label{24}
		\\
		u^{\prime}&=&\frac{1}{4}\log \frac{l_{v}^{2}\left( l_u+2u \right) ^2-4\left( l_uv+2\left( uv+z^2 \right) \right) ^2}{l_{v}^{2}\left( l_u-2u \right) ^2-4\left( l_uv-2\left( uv+z^2 \right) \right) ^2},
		\\
		v^{\prime}&=&\frac{1}{4}\log \frac{l_{u}^{2}\left( \left( -l_v \right) +2v \right) ^2-4\left( \left( -l_v \right) u+2\left( uv+z^2 \right) \right) ^2}{l_{u}^{2}\left( \left( -l_v \right) -2v \right) ^2-4\left( \left( -l_v \right) u-2\left( uv+z^2 \right) \right) ^2}.
	\end{eqnarray}
	Interestingly, this new transformation also turns the Poincar\'e AdS$_3$ metric into the form \eqref{7}. Correspondingly, in the boundary conformal field theory, this transformation becomes
	\begin{equation}
		\begin{aligned}
			u^{\prime}  = \operatorname{arctanh} \frac{2 u}{l_{u}}, ~~~~
			v^{\prime}  =-\operatorname{arctanh} \frac{2 v}{l_{v}} .\label{35}
		\end{aligned}
	\end{equation}
	This transformation can map the domain of dependecne $\mathcal{D}$ to $\mathcal{B}$ as well. Therefore, the thermal entropy in $\mathcal{B}$ remains unchanged and can also be given by the horizon entropy of \eqref{7}. Then, we substitute the horizon $z^\prime_h=1$ into Eq.\eqref{24}, and obtain
	\begin{equation}
		u=-\frac{1}{2}\sqrt{l_{u}^{2}+4\frac{l_u}{l_v}z^2},\qquad v=\frac{1}{2}\sqrt{l_{v}^{2}+4\frac{l_v}{l_u}z^2},\label{28}
	\end{equation}
	or
	\begin{equation}
		u=\frac{1}{2}\sqrt{l_{u}^{2}+4\frac{l_u}{l_v}z^2},\qquad v=-\frac{1}{2}\sqrt{l_{v}^{2}+4\frac{l_v}{l_u}z^2}.\label{29}
	\end{equation}
	These are exactly the two spacelike geodesics in the gravitational duality of the timelike entanglement entropy mentioned in the references \citep{Doi:2023zaf,Doi:2022iyj,Li:2022tsv}. To see this more clearly, we take $l_{u}=l_{v}=l$, then equations \eqref{28} and \eqref{29} become
	\begin{equation}
		u=-\frac{1}{2}\sqrt{l^2+4z^2},\qquad v=\frac{1}{2}\sqrt{l^2+4z^2}
	\end{equation}
	and
	\begin{equation}
		u=\frac{1}{2}\sqrt{l^2+4z^2},\qquad v=-\frac{1}{2}\sqrt{l^2+4z^2}.
	\end{equation}
	Note that $u=x+t$ and $v=x-t$, then we have 
	\begin{equation}\label{lz}
		x=0,\qquad t=\pm \frac{1}{2}\sqrt{l^2+z^2}.
	\end{equation}
	If we change the notation to be $l=T_{0}$, then Eq.\eqref{lz} are precisely the spacelike geodesics in Eq.\eqref{77}, and their areas are related to the real part of the timelike entanglement entropy. By connecting the endpoints of the two spacelike geodesics at null infinity with timelike geodesics, we can finally obtain the complete form of the holographic timelike entanglement entropy,  i.e., $S_{\rm TEE}=S_{\rm thermal}+ic\pi/6$.
	
	\section{Conclusion and discussion}\label{sec:con}

	In this paper, we re-examine the holographic entanglement entropy in AdS$_3$/CFT$_2$. For a Lorentz invariant theory, we believe that the entanglement entropy should be a function of the  domain of dependence of the region under consideration. By generalizing the concept of cut-off or defining a cut-off that is applicable to both spacelike and timelike regions, we reintroduce the timelike entanglement entropy. We find that the timelike entanglement entropy is the thermal entropy of the CFT after the Rindler transformation plus the constant $ic\pi/6$. Moreover, we obtained the gravitational dual of the timelike entanglement entropy, which are consistent with previous results.
	
	There are some interesting issues which need further investigations. For instance, in the usual Rindler approach, the Rindler transformation is a symmetric transformation, for instance in conformal field theory it is a conformal transformation. This symmetric transformation induces a unitary operator $U$ in Hilbert space, which relates the density matrix $\rho$ of the vacuum state in the original $u$, $v$ coordinate system to the thermal density matrix $\rho^\prime$ in the $u^\prime$, $v^\prime$ coordinate system,
	\begin{equation}
		\rho^\prime=U\rho U^\dagger.
	\end{equation}
	Since the unitary transformation does not change the von Neumann entropy, the entanglement entropy of the vacuum state is equal to the thermal entropy of the transformed thermal state \citep{Casini:2011kv}. 
	
	If we assume that the timelike entanglement entropy is also the von Neumann entropy of some ``density matrix", since the current Rindler transformation \eqref{35} is no longer a conformal transformation, we have reasons to believe that the timelike entanglement entropy is related to the ``transformed thermal entropy", but not exactly equal to it. This is in perfect agreement with equation \eqref{31} and also explains why we were unable to directly obtain the timelike geodesics \eqref{88} connecting the two spacelike geodesics from the Rindler method. Moreover, one can consider the new ``Rindler transformation" being composed of a transformation 
	\begin{equation}
		u^{\prime\prime}=u,\qquad v^{\prime\prime}=-v\label{43}
	\end{equation}
	and original Rindler transformation
	\begin{equation}
		u^{\prime}=\mathrm{arc}\tanh \frac{2u^{\prime\prime}}{l_u},\qquad v^{\prime}=\mathrm{arc}\tanh \frac{2v^{\prime\prime}}{l_v}.
	\end{equation}
	Transformation \eqref{43} makes $ds^2\rightarrow-ds^2$. Such a transformation will turn timelike curves into spacelike curves and vice versa. For convenience, this paper refers to it as the ``dual transformation". Although we do not know what properties the dual transformation has, we can naively conjecture that under the dual transformation, the entanglement entropy remains unchanged up to a constant $ic\pi/6$. (Further detailed discussion can be found in the Appendix \ref{B}.)
	
	Holographic entanglement entropy also leads to an interesting idea that the spatial coordinates of AdS come from quantum entanglement \citep{Swingle:2009bg,VanRaamsdonk:2010pw}. It is natural to think about whether the time coordinate also comes from quantum entanglement. Timelike entanglement entropy is likely to be a quantity related to the emergence of the time coordinate, although its physical meaning is not yet clear \citep{Doi:2023zaf}. However, it opens a new window for generalizing entanglement entropy and deserves further studies. 
	
	\section*{Acknowledgement} This work was partially supported by the National Natural Science Foundation of China (Grants No.12175008).
	
	\appendix
	\section{Define timelike entanglement entropy via Wick rotation}\label{A}
	Consider a free scalar field on a cylinder, with mass $m$. The space and time coordinates are denoted by $x$ and $t$, respectively. Assuming the spatial circle is $x\sim x+R$. We can easily write down the action of the scalar field
	\begin{equation}
		Z_{\phi}=\int D \phi e^{i S}.
	\end{equation}
	To define the timelike entanglement entropy, we treat $t$ as the spatial direction and $x$ as the Euclidean time. It can be considered that spacetime has rotated by ninety degrees. Then, the ``timelike Hamiltonian" $H$ is
	\begin{equation}
		H=-\frac{i}{2} \int d t\left[\pi^{2}+\left(\partial_{t} \phi\right)^{2}-m^{2} \phi^{2}\right],
	\end{equation}
	where
	\begin{equation}
		\pi=-\partial_x\phi.
	\end{equation}
	Then, the partition function is
	\begin{equation}
		Z_{\phi}=\operatorname{Tr}\left[e^{-R H}\right].\label{A.4}
	\end{equation}
	We introduce the rescaled Hamiltonian $\tilde{H} = iH$, which takes the conventional form with a minus sign for the mass term
	\begin{equation}
		\tilde{H}=\frac{1}{2} \int d t\left[\pi^{2}+\left(\partial_{t} \phi\right)^{2}-m^{2} \phi^{2}\right].
	\end{equation}
	\eqref{A.4} can be rewritten as 
	\begin{equation}
		Z_{\phi}=\operatorname{Tr}\left[e^{i R \tilde{H}}\right]=\text{Tr}\left [e^{-\beta_s\tilde{H}}\right ].
	\end{equation}
	$1/\beta_s$ is the temperature of the cylinder after spacetime rotation. A spacelike region $A$ on this cylinder corresponds to a timelike region on the cylinder before rotation. We can obtain the timelike entanglement entropy by making the following substitution to the usual entanglement entropy formula:
	\begin{equation}
		\beta_{S} \rightarrow-i R, \quad m \rightarrow-i m .\label{A.7}
	\end{equation}
	For two-dimentional CFT, the entanglement entropy for the thermal state at temperature $ 1/\beta_s$ is
	\begin{equation}
		S_{A}=\frac{c}{3} \log \left[\frac{\beta_{S}}{\pi \tilde{\epsilon}} \sinh \frac{\pi X}{\beta_{S}}\right],
	\end{equation}
	where $X$ is the length of the spatial interval $A$. $\tilde{\epsilon}$ is the cut-off defined for $\tilde{H}$, while $\varepsilon=i\tilde{\epsilon}$ is defined for the Hamiltonian $H$ we are interested in. By making the substitution \eqref{A.7} as mentioned earlier and taking $X=T_0$, we obtain the timelike entanglement entropy
	\begin{equation}
		S_{A}^{(t)}=\frac{c}{3} \log \left[\frac{R}{\pi \epsilon} \sin \frac{\pi T_{0}}{R}\right]+\frac{i \pi c}{6} .
	\end{equation}
	If we let $R\rightarrow\infty$, we obtain 
	\begin{equation}
		S_{A}^{(t)}=\frac{c}{3} \log \frac{T_{0}}{\epsilon}+\frac{c \pi}{6} i .
	\end{equation}
	This is precisely \eqref{2.3} in the case of a pure timelike interval.
	
	\section{The origin of $ic\pi/6$}\label{B}
	
	To express our ideas more clearly, we would like to provide some explanations of the origins of $ic\pi/6$. In the preceding text, there is a coordinate transformation
	\begin{equation}
		u^{\prime\prime}=u,\qquad v^{\prime\prime}=-v.
	\end{equation}
	This transformation is equivalent to swapping the spacetime coordinates
	\begin{equation}
		x^{\prime\prime}=t,\qquad t^{\prime\prime}=x.
	\end{equation}
	Then we have
	\begin{equation}
		ds^{2}=-dt^{2}+dx^{2},\qquad ds^{\prime\prime 2}=-dt^{\prime\prime 2}+dx^{\prime\prime 2}
	\end{equation}
	and
	\begin{equation}
		ds^{2}=-ds^{\prime\prime 2}.
	\end{equation}
	Suppose $ds^{2}$ and $ds^{\prime\prime 2}$ describe the geometries $M$ and $M^{\prime\prime}$, respectively. Then a timelike curve in $M$ corresponding to a spatial curve in $M^{\prime\prime}$. In $M^{\prime\prime}$, we can use ordinary Rindler transformation to get entanglement entropy of the spatial curve, which has the following form
	\begin{equation}
		S=S(\text{Domain of dependence},\text{cutoffs}),
	\end{equation}
	in which the cutoffs should be calculated by the spacetime interval. We want to define the timelike entanglement entropy in $M$ through the entanglement entropy of $M^{\prime\prime}$ by directly replacing the domain of dependence and cutoffs in $S$. Interestingly, the corresponding domains of dependence on $M$ and $M^{\prime\prime}$ are identical, but the cutoffs differ, leading to a term $ic\pi/6$ in the timelike entanglement entropy.

	
	\bibliography{references}

\end{document}